\documentstyle[aps,prd,preprint,tighten,epsf]{revtex}
\tighten
\begin{document}
\preprint{		\makebox{\begin{tabular}{r}
                        				INFNFE-02-98	\\
							BARI-TH/297-98	\\
                                                      	OUTP-98-14-P	\\
                                                      	astro-ph/9803177\\
                                                                        \\
                                 \end{tabular}}}
\draft
\title{Quantifying uncertainties in primordial nucleosynthesis \\ 
	without Monte Carlo simulations}
\author{G.\ Fiorentini~${}^a$, E.\ Lisi~${}^b$, S.\ Sarkar~${}^c$, and
         F.~L.\ Villante~${}^a$ \\}
\address{$^a$Dipartimento di Fisica and Sezione INFN di Ferrara, 
             Via del Paradiso 12, I-44100 Ferrara, Italy \\
         $^b$Dipartimento di Fisica and Sezione INFN di Bari,
             Via Amendola 173, I-70126 Bari, Italy \\
         $^c$Theoretical Physics, University of Oxford, 
             1 Keble Road, Oxford OX1 3NP, UK }
\maketitle
\bigskip
\begin{abstract}
We present a simple method for determining the (correlated)
uncertainties of the light element abundances expected from big bang
nucleosynthesis, which avoids the need for lengthy Monte Carlo
simulations. Our approach helps to clarify the role of the different
nuclear reactions contributing to a particular elemental abundance and
makes it easy to implement energy-independent changes in the measured
reaction rates. As an application, we demonstrate how this method
simplifies the statistical estimation of the nucleon-to-photon ratio
through comparison of the standard BBN predictions with the
observationally inferred abundances.
\end{abstract}
\pacs{26.35.+c, 98.80.Ft}

\section{Introduction}

	Big bang nucleosynthesis is entering the precision era
\cite{Sc97}. On the one hand, there has been major progress in the
observational determination of the abundances of the light elements D
\cite{Bu97,We97}, $^3$He \cite{Ba94,Gl96}, $^4$He \cite{Ol97a,Iz97},
and $^7$Li\cite{Ry96,Bo97}, although the increasing precision has
highlighted discrepancies between different measurements (see
Refs.\cite{Mo97,Ho97,Le97} for recent assessments). Secondly, we have
a sound analytical understanding of the physical processes involved
\cite{Be89,Es91} and the standard BBN computer code \cite{Wa73,Ka92}
which incorporates this physics is robust and can be easily altered to
accomodate changes in the input parameters, e.g.\ nuclear reaction
rates \cite{Sm93}. The comparison of increasingly accurate
observationally inferred and theoretical abundances will further
constrain the values of fundamental parameters, such as the nucleon
density parameter (see, e.g., Ref.\cite{Ol97b}) or extra degrees of
freedom related to possible new physics beyond the Standard Model
(see, e.g., Ref.\cite{Sa96}). It goes without saying that error
evaluation represents an essential part of such comparisons.

	Because of the complex interplay between different nuclear
reactions, it is not straightforward to assess the effect on a
particular elemental yield of the uncertainties in the experimentally
determined reaction rates. The authors of Ref.\cite{Kr90} first
employed Monte Carlo methods to sample the error distributions of the
relevant reaction cross-sections which were then used as inputs to the
standard BBN computer code. This enables well-defined confidence
levels to be attached to the theoretically predicted abundances,
e.g. the abundance range within which say 95\% of the computed values
fall correspond to 95\% C.L. limits on the expected abundance. It was
later realized that error correlations are also relevant, and can be
estimated with the same technique \cite{Ke94,KeKr}. The Monte Carlo
(MC) approach has since become the standard tool for comparing theory
and data \cite{Sm93,Kr94,Co95,Ha95,Ol97c}. However, although it can
include refinements such as asymmetric or temperature-dependent
reaction rate uncertainties \cite{Sm93}, it requires lengthy
calculations which need to be repeated each time (any of) the input
parameters are changed or updated. Since we may expect continued
improvement in the determination of the relevant parameters, it is
desirable to have a faster method for error evaluation and comparison
with observations.

	In this work we propose a simple method for estimation of the
BBN abundance uncertainties and their correlations which requires
little computational effort. The method, based on linear error
propagation, is described in Sec.~II. A concrete application is given
in Sec.~III, where theory and observations are compared using simple
$\chi^2$ statistics to obtain the best-fit value of the
nucleon-to-photon ratio. In Sec.~IV we study with this method the
relative importance of different nuclear reactions in determining the
synthesized abundances. Conclusions and perspectives for further work
are presented in Sec.~V.

\section{Propagating input cross section uncertainties to output
          elemental abundances}

\subsection{Notation and input}

	The four relevant element abundances $Y_i$ considered in this
work are defined in Table~I. (Note that the abundance of $^4$He is
conventionally quoted as a {\em mass fraction}, while the abundances
of D, $^3$He and $^7$Li are ratios by number.) In BBN calculations,
the $Y_i$'s depend both on model parameters (the nucleon-to-photon
ratio $\eta$, the number of neutrino families $N_\nu$, etc.) and on a
network of nuclear reactions $R_k$:
\begin{equation}
		Y_i = Y_i (\eta,\,N_\nu,\,\ldots;\,\lbrace R_k\rbrace)\ .
\end{equation}

	The most important $R_k$'s are listed, numbered as in
Ref.\cite{Ka92}, in the first two columns of Table~II, while our
default inputs for the rates $R_k$ and their $1\sigma$ uncertainties
$\pm\Delta\,R_k$ are given in the third and fourth columns. The
numerical values are given in ratio to the reference reaction rates
compiled in Table~1 of Ref.\cite{Sm93}; we have chosen identical
values (i.e., $R_k=1$) except for $R_1$, the neutron decay rate, where
we adopt the most recent world average for the neutron lifetime of
$\tau_n=886.7\pm1.9$~s \cite{taun,tau1}, as compared to the value of
$\tau_n=888.54\pm3.73$~s used in Ref.\cite{Sm93}. The fractional
uncertainties $\pm\Delta R_k/R_k$ $(k\neq1)$ have also been taken from
Ref.\cite{Sm93} (see their Table~2), assuming conservatively the
largest value for the temperature-dependent errors $\Delta\,R_7$ and
$\Delta\,R_{10}$.%
\footnote{Our method for error propagation requires that the
 $\Delta\,R_k/R_k$'s be constant (i.e., temperature independent). We will
 comment on this point at the end of this Section.}

\subsection{The method}
   
	For simplicity we consider only the standard BBN case (i.e.,
$N_\nu=3$, etc.), so that $\eta$ is the only model parameter being
varied in the calculation of the abundances $Y_i$ and of their
uncertainties $\sigma_i$:
\begin{equation}
		Y_i = Y_i(\eta)\pm \sigma_i(\eta)\ .
\label{Yi}
\end{equation}
Our method can however be easily generalized to nonstandard cases.

	For a relatively small change $\delta\,R_k$ of the input rate
$R_k$ $(R_k\to R_k+\delta\,R_k)$, the corresponding deviation
$\delta\,Y_i$ of the $i$-th elemental abundance $(Y_i\to
Y_i+\delta\,Y_i)$, as given by linear propagation, reads
\begin{equation}
	\delta\,Y_i (\eta) = Y_i(\eta) \sum_k 
	\lambda_{ik}(\eta)\frac{\delta\,R_k}{R_k}\ ,
\label{dYi}
\end{equation}
where the functions $\lambda_{ik}(\eta)$ represent the logarithmic
derivatives of $Y_i$ with respect to $R_k$:
\begin{equation}
	\lambda_{ik}(\eta) = 
		\frac{\partial\ln Y_i(\eta)}{\partial\ln R_k(\eta)}\ .
\label{lambdaik}
\end{equation}

	In general, the deviations $\delta\,Y_i$ in Eq.~(\ref{dYi})
are correlated, since they all originate from the same set of reaction
rate shifts $\{\delta\,R_k\}$. The global information is contained in
the error matrix (also called covariance matrix) \cite{stat}, which is
a generalization of the ``error vector'' $\delta\,Y_i$ in
Eq.~(\ref{dYi}).  In particular, the abundance error matrix
$\sigma^2_{ij}(\eta)$ obtained by linearly propagating the input
$\pm1\sigma$ reaction rate uncertainties $\pm\Delta\,R_k$ to the
output abundances $Y_i$ reads:
\begin{equation}	
	\sigma^2_{ij}(\eta) = Y_i(\eta) Y_j(\eta) \sum_k
	\lambda_{ik}(\eta)\lambda_{jk}(\eta) 
	\left(\frac{\Delta\,R_k}{R_k}\right)^2\ .
\label{sigmaij}
\end{equation}
This matrix completely defines the abundance uncertainties. In
particular, the $1\sigma$ abundance errors $\sigma_i$ of
Eq.~(\ref{Yi}) are given by the square roots of the diagonal elements,
\begin{equation}
	\sigma_i(\eta)=\sqrt{\sigma^2_{ii}(\eta)}\ ,
\label{sigmaii}
\end{equation}
while the error correlations  $\rho_{ij}$ can be derived from 
Eqs.~(\ref{sigmaij},\ref{sigmaii}) through the standard definition
\begin{equation}
	\rho_{ij}(\eta) = 
		\frac{\sigma_{ij}^2(\eta)}{\sigma_i(\eta)\,\sigma_j(\eta)}\ .
\label{rhoij}
\end{equation}
Thus Eqs.(\ref{Yi}--\ref{rhoij}) represent all that is required to
calculate the errors in the predicted abundances and their
correlations.

	Note that the relevant physics is contained entirely in the
central values $Y_i$ and in their logarithmic derivatives
$\lambda_{ik}$, which have to be evaluated just once with a BBN
numerical code, thus dramatically reducing the required computing
time.
\footnote{The logarithmic derivatives are numerically defined as
 $\lambda_{ik}(\eta)=
 [Y_i(\eta,\,R_k+\Delta\,R_k)-Y_i(\eta,\,R_k-\Delta\,R_k)]R_k
 /2Y_i(\eta,\,R_k)\Delta\,R_k$, at given $\eta$. We find negligible
 difference between left and right derivatives.}
We have made a further check of the linearity of the error propagation
by calculating the logarithmic derivatives with increments equal to
$\Delta\,R_{k}$ (default) and $2\,\Delta\,R_{k}$, obtaining
practically the same functions $\lambda_{ik}$ in either case. This
means that doubling the error on $R_k$ also doubles the corresponding
error component of $Y_i$, i.e. the error propagation is indeed linear.

\begin{figure}[tbh]
\epsfysize20truecm
\vskip1cm
\epsfbox{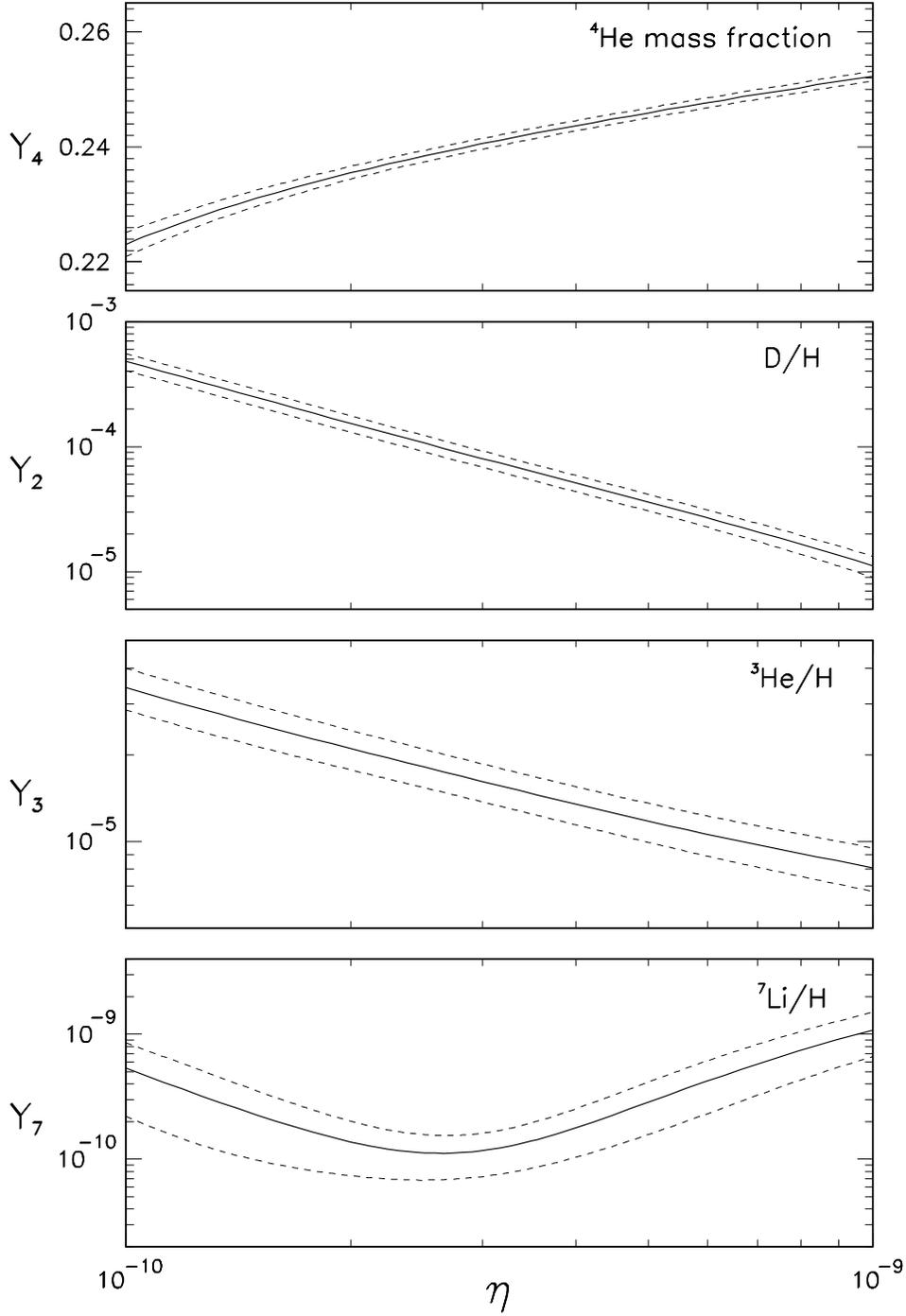}
\caption{Primordial abundances $Y_i$ (solid lines) and their $\pm
 2\sigma$ bands (dashed lines), as functions of the nucleon-to-photon
 ratio $\eta$.}
\label{fig1}
\end{figure}

\begin{figure}[tbh]
\epsfysize21truecm
\epsfbox{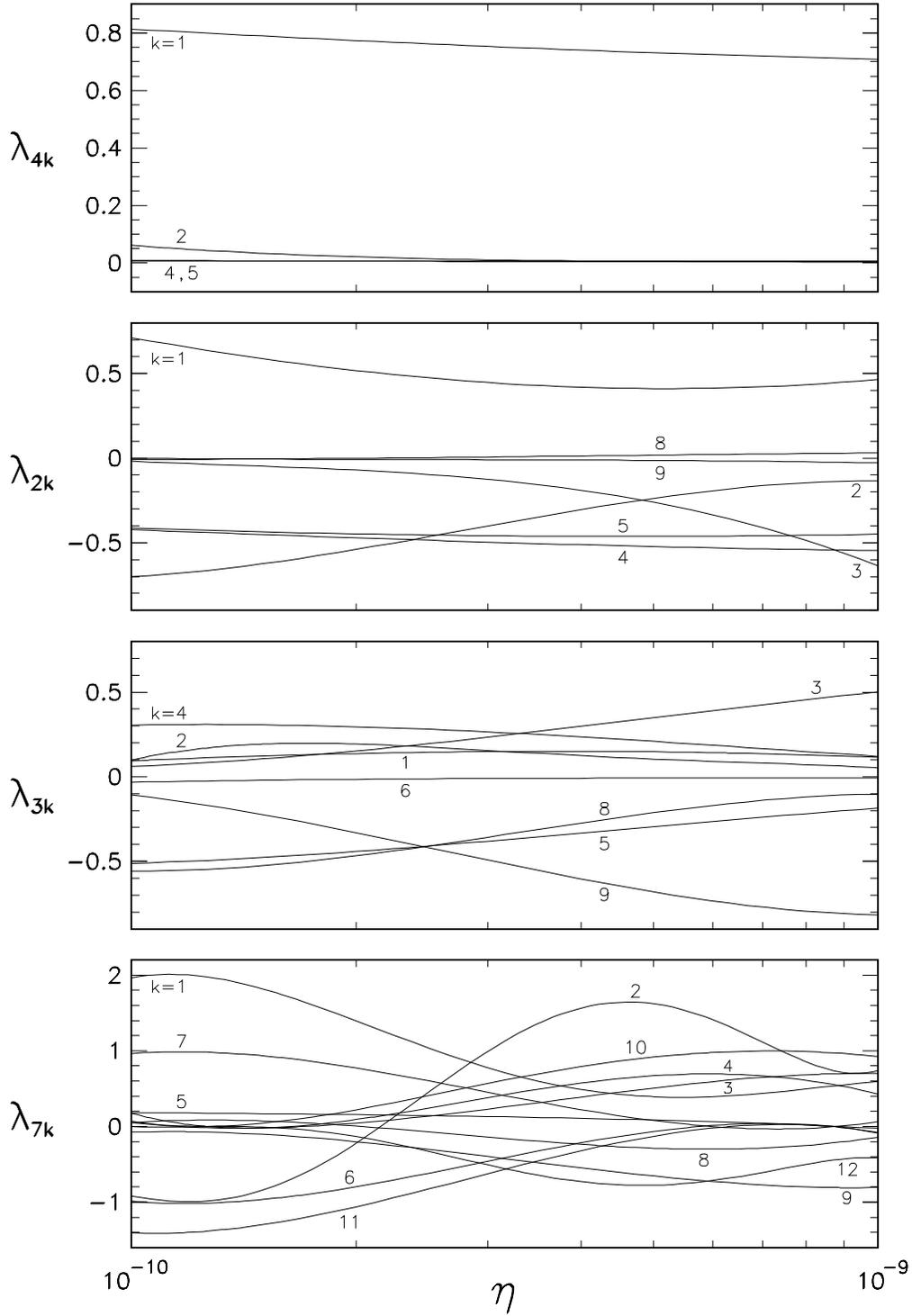}
\caption{Logarithmic derivatives $\lambda_{ik}$ of the abundances
 $Y_i$ with respect to the reaction rates $R_k$, as functions of $\eta$.}
\label{fig2}
\end{figure}

\begin{figure}[tbh]
\epsfysize21truecm
\epsfbox{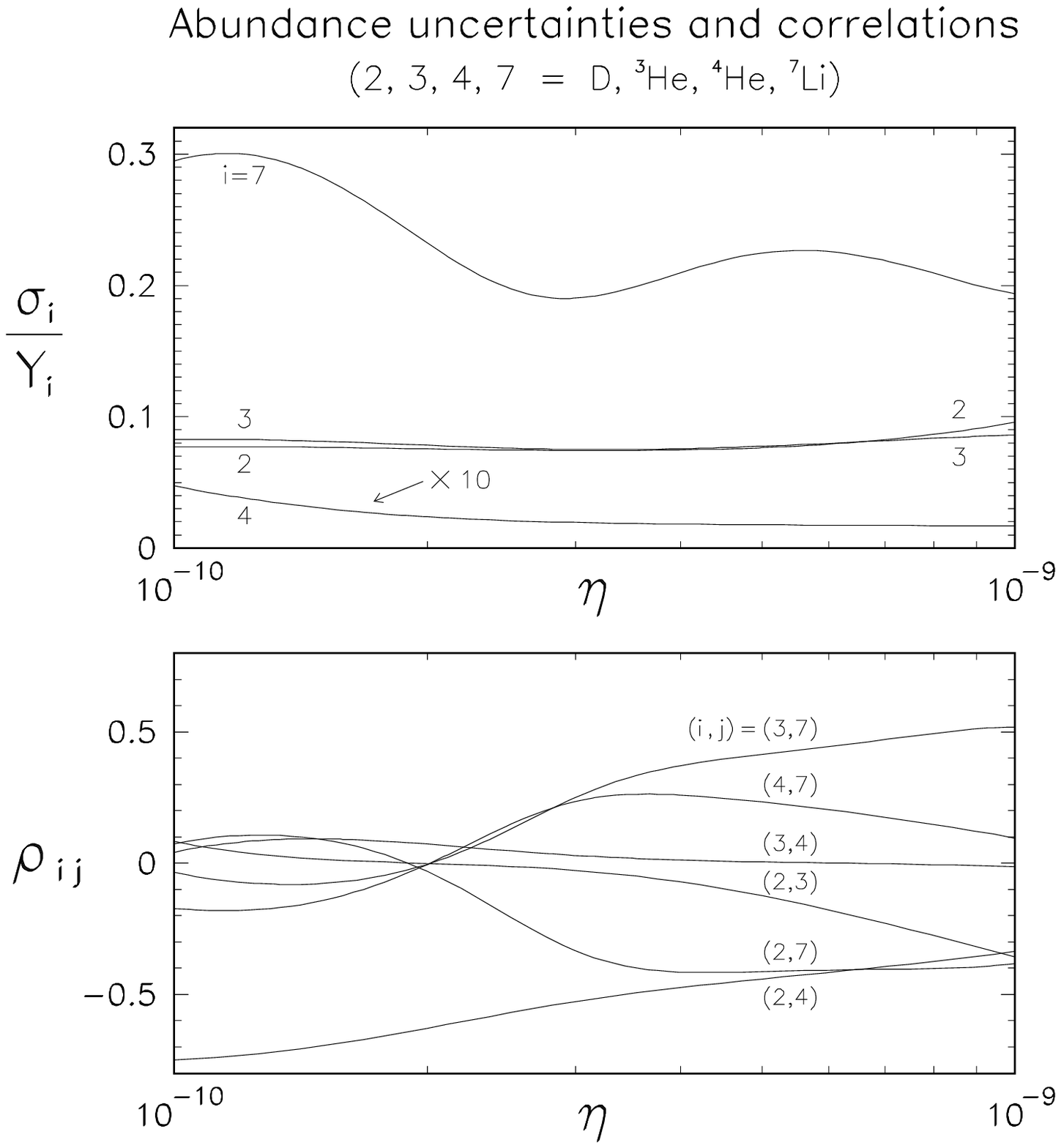}
\caption{Fractional abundance uncertainties $\sigma_i/Y_i$ (upper panel)
 and their correlations $\rho_{ij}$ (lower panel), as functions of $\eta$. }
 \label{fig3}
\end{figure}

	We think it useful to present the results of this exercise in
the form of tables so all calculations that will now follow can be
done on a pocket calculator. Tables~III and IV show the coefficients
of polynomial fits to $Y_i$ and $\lambda_{ik}$, respectively, for
$\eta$ in the usually considered range $10^{-10}-10^{-9}$.%
\footnote{Small corrections to the helium abundance $Y_4$ due to
 Coulomb, radiative and finite temperature effects, finite nucleon
 mass effects and differential neutrino heating, have been
 incorporated according to the prescription given in
 Ref.~\protect\cite{Sa96}. We have used the BBN code 
 \protect\cite{Ka92} with the
 lowest possible settings of the time steps in the (2nd order)
 Runge-Kutta routine, which allows rapid convergence to within
 $0.01\%$ of the true value \protect\cite{Ke93}. We understand that our
 results are in good agreement with a recent independent computation
 of $Y_4$ using a new BBN computer code \protect\cite{Lo98}.}
The abundances $Y_i(\eta)$ with their associated $\pm2$ standard
deviation error bands calculated through Eq.~(\ref{sigmaii}) are shown
in Fig.~\ref{fig1}. The functions $\lambda_{ik}(\eta)$ are shown in
Fig.~\ref{fig2}; note that some of these vary strongly (and even
change sign) with $\eta$, indicating that the physical dependence of
the $Y_i$'s on the $R_k$'s may be quite subtle. Although some general
features of this dependence have been addressed in Ref.\cite{Es91},
further work is needed to interpret the functional form of the
$\lambda_{ik}$'s in Fig.~\ref{fig2}. We intend to address this issue
elsewhere. Finally, Fig.~\ref{fig3} displays the fractional
uncertainties $\sigma_i/Y_i$ and their correlations $\rho_{ij}$ as
derived from Eqs.~(\ref{sigmaij}--\ref{rhoij}). Notice that, in
general, the error correlations are non-negligible and should be
properly taken into account in statistical analyses, as first
emphasized in Ref.\cite{Ke94}.

	In summary, the recipe for evaluating the BBN uncertainties
$\pm \sigma_i$ affecting the $Y_i$'s for a given value of $\eta$ is:

\begin{description}

\item[\ \ ($i$)] Determine the abundances $Y_i(\eta)$ and their
	logarithmic derivatives $\lambda_{ik}(\eta)$ using Tables III
	and IV, respectively;

\item[\ ($ii$)] If the central values of the reaction rates $R_k$ are
	updated $(R_k\to R_k+\delta\,R_k)$ with respect to those reported
	in Table~II, then update also the central values of the
	abundances $(Y_i\to Y_i +\delta\,Y_i)$ through
	Eq.~(\ref{dYi});

\item[($iii$)] For given reaction rate uncertainties $\Delta\,R_k$
	(e.g., from Table~III), compute the abundance errors
	$\sigma_i$ and their correlations $\rho_{ij}$ using
	Eqs.~(\ref{sigmaij}--\ref{rhoij}).

\end{description}


\subsection{Comparison with MC estimates and remarks}

	Our approach is based on the linear propagation of errors
originating from many independent sources (i.e. the $R_k$'s). One can
expect that this method will work reasonably well, both because the
input fractional uncertainties $\Delta\,R_k/R_k$ are relatively small,
and because the final output uncertainties $\sigma_i$ affecting the
abundances $Y_i$ are ``regularized'' by the central limit
theorem. Indeed, our $\pm2\sigma$ bands in Fig.~\ref{fig1} compare
well with the MC-estimated bands of
Refs.~\cite{Sm93,Ke94,Kr94,Co95,Ha95,Ol97c}, with small relative
differences which depend, in part on different input $R_k\pm
\Delta\,R_k$'s, and that are not larger than the spread among the
various MC estimates themselves.

\begin{figure}[tbh]
\epsfysize21truecm
\epsfbox{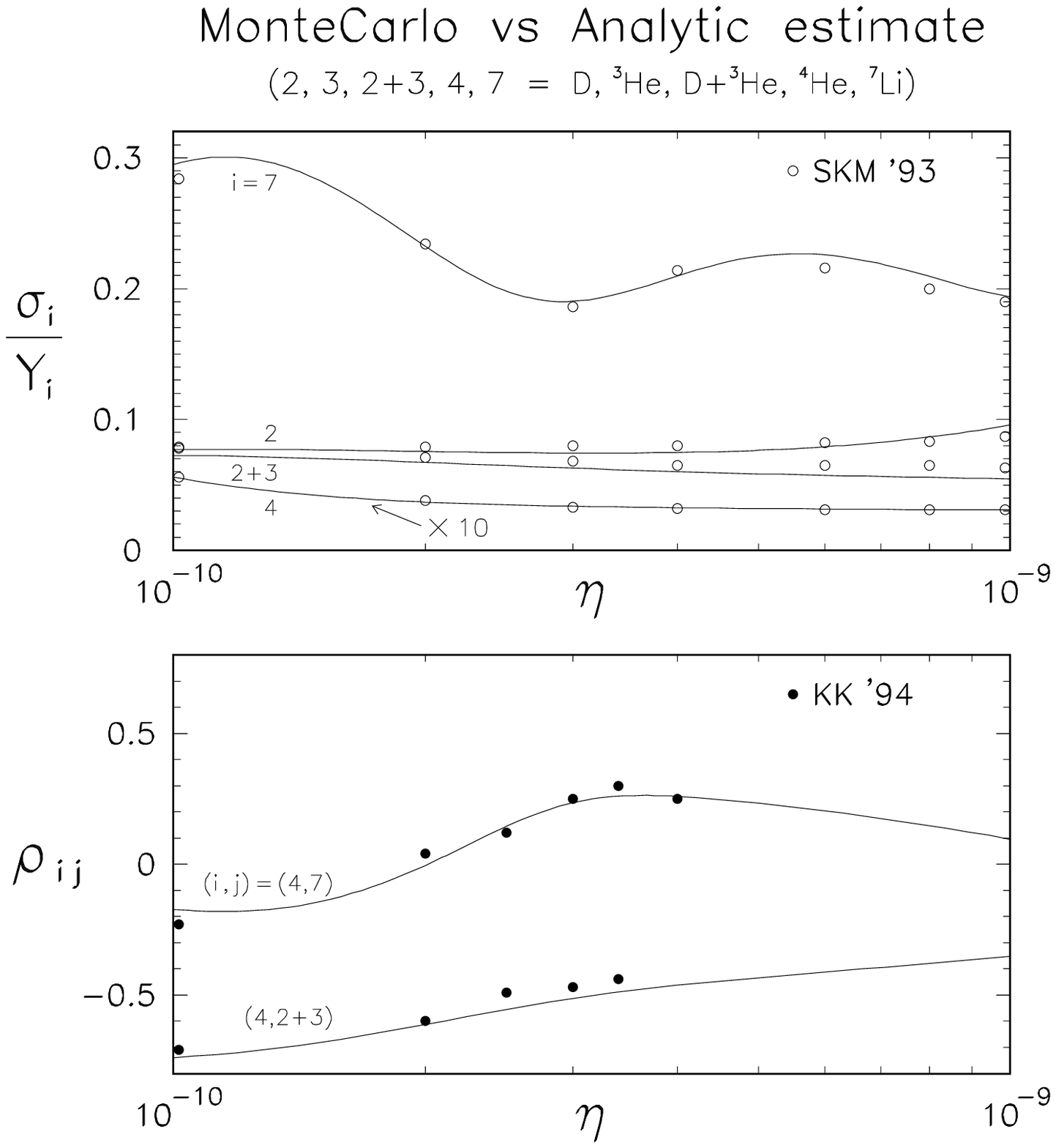}
\caption{Monte Carlo estimates of $\sigma_i/Y_i$ (SKM~'93
 \protect\cite{Sm93}, dots) and $\rho_{ij}$ (KK~'94
 \protect\cite{KeKr}, dots), compared with our analytic evaluation
 (solid lines), using the same inputs.}
\label{fig4}
\end{figure}

	In order to be more quantitative, we compare in
Fig.~\ref{fig4} (upper panel) the MC evaluation of the fractional
uncertainties $\sigma_i/Y_i$ as derived from Ref.\cite{Sm93}%
\footnote{The MC values of $\sigma_i/Y_i$ have been read off the
 (small) panels of Fig.~27 in Ref.\protect\cite{Sm93} and, therefore,
 may be subject to small transcription errors.}
with our analytic estimate (using, for this exercise, the same input
parameters). There is good agreement between these two totally
independent estimates. In Fig.~\ref{fig4} (lower panel) we also show a
comparison with the only MC evaluation of $\rho_{ij}$ we are aware of
(viz., Ref.\cite{KeKr}), obtaining again good agreement with our
calculation when the same input $R_k\pm \Delta\,R_k$ are used. We
conclude the discussion of Fig.~\ref{fig4} by noting that the
uncertainty $\sigma_{(2+3)}$ and the correlations $\rho_{(2+3)j}$
related to the often-used combination of abundances
$Y_{(2+3)}=Y_2+Y_3=({\rm D}+{}^3{\rm He})/{\rm H}$ are given, within
our approach, by:
\begin{eqnarray}
\sigma^2_{(2+3)} &=& \sigma_2^2 + \sigma_3^2 + 
 2\rho_{23}^{\phantom{2}} \sigma_2^{\phantom{2}} \sigma_3^{\phantom{2}}\ , \\
\rho_{(2+3)j}^{\phantom{2}}\sigma_{(2+3)}^{\phantom{2}} \sigma_j^{\phantom{2}}
 &=& \rho_{2j}^{\phantom{2}}\sigma_2^{\phantom{2}} \sigma_j^{\phantom{2}} +
 \rho_{3j}^{\phantom{2}}\sigma_3^{\phantom{2}} \sigma_j^{\phantom{2}}\ .
\end{eqnarray}

	There are, of course, some refined features of the MC approach
that cannot be addressed with our method, such as asymmetric or
temperature-dependent uncertainties $\Delta\,R_k$ \cite{Sm93}. However
we consider these refinements not essential for practical
applications.  In a sense, the possible asymmetry between ``upper''
and ``lower'' errors is where one wants it to be. For instance, if one
assumes {\em a priori\/} symmetric errors in the astrophysical
$S$-factors, then asymmetric errors are induced in the thermally
averaged reaction rates $R\sim \langle \sigma v\rangle$; conversely,
the requirement of {\em a priori\/} symmetric $\Delta\,R_k$ errors
requires that the input $S$-factor uncertainties are readjusted, as
discussed in Ref.\cite{Sm93}. Although the authors of Ref.\cite{Sm93}
have adopted the latter option $(|+\Delta\,R_k|=|-\Delta\,R_k|)$, the
former option or others are equally acceptable, and would clearly
produce different outputs for the MC estimate of the abundance
errors. For instance, the upper and lower errors of $Y_7$ appear to be
rather symmetrical in the MC calculation of Ref.\cite{Sm93}, while
they are noticeably asymmetrical in Ref.\cite{KeKr}.

	Concerning the temperature-dependent \cite{Sm93} uncertainties
$\Delta\,R_{7}$ and $\Delta\,R_{10}$, which affect mainly the estimate
of $Y_7$, our conservative choice in Table~II proves to be successful
for the estimate of $\sigma_7/Y_7$ (see Fig.~\ref{fig4}). This seems to
indicate that the uncertainties at low temperatures (which are larger
than those at high temperatures \cite{Sm93}) dominate in the estimate
of these errors.  In any case, since practically all reaction rate
uncertainties $\Delta\,R_k$ contribute to the final value of
$\sigma_7$, temperature-dependent refinements in the propagation of
just two of these ($\Delta\,R_7$ and $\Delta\,R_{10}$) do not appear
to be decisive for the estimate of the global error $\sigma_7$.

	In conclusion, we have shown that our simple analytic method
for error evaluation represents an useful alternative to lengthy and
computationally expensive MC simulations. Both the magnitude and the
correlations of the total errors affecting the theoretical abundances
are reproduced with good accuracy. We therefore advocate the use of
this method for BBN analyses as an alternative to MC simulations.%
\footnote{ A parallel situation holds in the field of solar neutrino
 physics, where the correlated uncertainties of the neutrino fluxes
 predicted by solar models have been estimated through both Monte
 Carlo simulations \cite{Ba88} and linear propagation of input errors
 \cite{Fo95}. The latter technique has proved more popular because of
 its ease of use.}


\section{Determining the likely nucleon-to-photon ratio}

	The comparison of the predicted primordial abundances
$Y_i(\eta)\pm\sigma_i(\eta)$ with their observationally inferred
values $\overline Y_i\pm\overline\sigma_i$ through a statistical test
allows extraction of the likelihood range for the fundamental
parameter $\eta$. So far, this has been done either through fit-by-eye
(see, e.g., Ref.\cite{Sm93}) or by Monte Carlo-based maximum
likelihood methods (see, e.g., Refs.\cite{Ha95,Ol97c}). In this
Section we show how limits on $\eta$ can be simply extracted using
$\chi^2$ statistics based on the method described in the previous
Section.

	Assuming that the errors $\overline\sigma_i$ in the
determinations of different abundances $\overline Y_i$ are
uncorrelated, the experimental squared error matrix
$\overline\sigma^2_{ij}$ is simply
\begin{equation}
\overline \sigma^2_{ij} = \delta_{ij}\overline\sigma_i\overline\sigma_j\ ,
\label{sigmaijexp}
\end{equation}
where $\delta_{ij}$ is Kronecker's delta. The total (experimental +
theoretical) error matrix $S^2_{ij}$ is then obtained by summing the
matrices in Eqs.~(\ref{sigmaij},\ref{sigmaijexp}):
\begin{equation}
S^2_{ij}(\eta) = \sigma^2_{ij}(\eta) + \overline\sigma^2_{ij}\ .
\end{equation}
Its inverse defines the weight matrix $W_{ij}(\eta)$:
\begin{equation}
W_{ij}(\eta) = [S^2_{ij}(\eta)]^{-1}.
\label{W}
\end{equation}
The $\chi^2$ statistic associated with the difference between
theoretical $(Y_i)$ and observational $(\overline{Y_i})$ light element
abundance determinations is then \cite{stat}:
\begin{equation}
\chi^2(\eta) = \sum_{ij} [Y_i(\eta)-\overline{Y_i}]\cdot W_{ij}(\eta)\cdot
[Y_j(\eta)-\overline{Y_j}]\ .
\label{chi}
\end{equation}
Minimization of the $\chi^2$ gives the most probable value of $\eta$,
while the intervals defined by $\chi^2=\chi^2_{\rm
min}+\Delta\,\chi^2$ give the likely ranges of $\eta$ at the
confidence level set by $\Delta\,\chi^2$ (for one degree of freedom,
$\eta$).%
\footnote{We remind the reader that $\Delta\,\chi^2 =
 1,\,2.71,\,3.84,$ and 6.64 correspond to confidence levels (C.L.'s)
 of $68\%,\,90\%,\,95\%,$ and 99\%, respectively.}

	In order to illustrate this, we estimate $\eta$ using recent
observational data for the three light element abundances
($Y_2,Y_4,Y_7)$ whose primordial origin is most secure. It is well
known that the observationally inferred values $\overline{Y_2}$ and
$\overline{Y_4}$ are still controversial, and the conflict between
different determinations has driven a lively debate on the status of
BBN (see Refs.\protect\cite{Ol97b,Sa97} and references therein). In
this paper we do not enter into this debate but rather apply our
method to two possible (although mutually incompatible) selections of
measurements which we name data set ``A'' and data set ``B'':
\begin{equation}
{\rm Data\  set\  A:\ }\left\{
\begin{array}{l}
\overline{Y_2} = 1.9 \pm 0.4 \times 10^{-4}  \ , \\
\overline{Y_4} =  0.234 \pm 0.0054             \ , \\
\overline{Y_7} =  1.6 \pm 0.36 \times 10^{-10}\ ; 
\end{array}\right.
\label{datasetA}
\end{equation}
\begin{equation}
{\rm Data\  set\  B:\ }\left\{
\begin{array}{l}
\overline{Y_2} =  3.40 \pm 0.25 \times 10^{-5}  \ , \\
\overline{Y_4} =  0.243 \pm 0.003             \ , \\
\overline{Y_7} =  1.73 \pm 0.12 \times 10^{-10}\ . 
\end{array}\right.
\label{datasetB}
\end{equation}

	The data set ``A'' is used in Ref.\cite{Ol97c}, the authors of
which make a detailed MC-based fit to $\eta$, thus enabling comparison
with our method. Note that their adopted value of the primordial
deuterium abundance from observations of high redshift quasar
absorption systems is consistent with another recent observation,
$\overline{Y_2}=(2.15\pm0.35)\times 10^{-4}$ \cite{We97}, but in
conflict with the significantly smaller value reported in
Ref.\cite{Bu97}, which we adopt for data set ``B''. Similarly the
primordial helium mass fraction inferred from observations of
metal-poor blue compact galaxies which we adopt for data set ``B'' is
from Ref.\cite{Iz97} in which it is argued that previous analyses
leading to the smaller value of $\overline{Y_2}$ used in data set
``A'' underestimate the true abundance (although this is disputed in
Ref.\cite{Ol97a}.) Finally the estimates for the primordial lithium
abundance in both data sets are based on observations of pop~II stars,
with the slightly higher value \cite{Bo97} of $\overline{Y_7}$ in data
set ``B'' taken from an updated analysis. Readers who prefer to adopt
different combinations of these, or indeed other, estimates for the
primordial abundances are invited to perform their own fit to $\eta$
by following the simple prescription given here.

\begin{figure}[tbh]
\epsfysize20truecm
\epsfbox{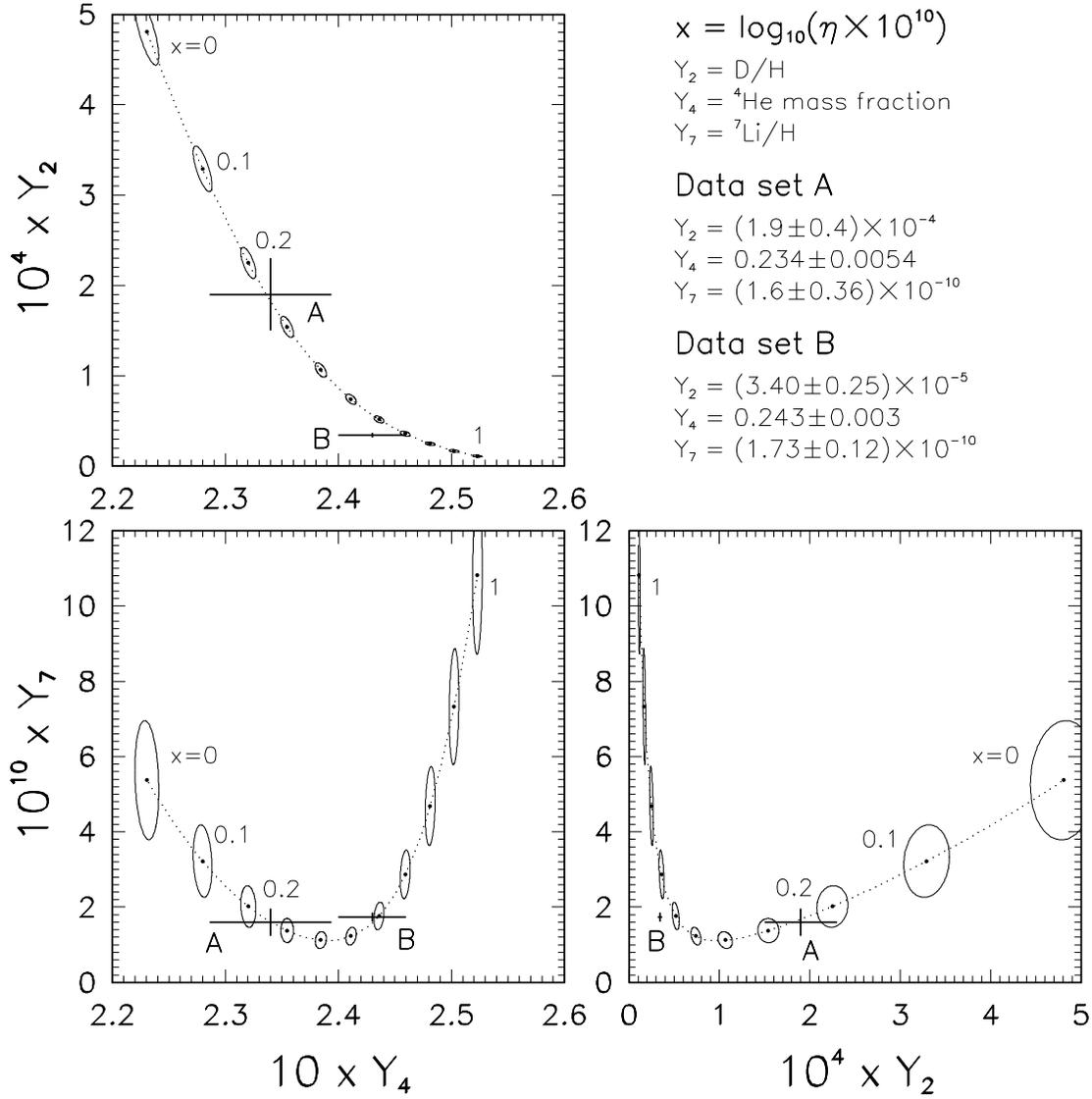}
\caption{Standard BBN predictions (dotted lines) in the 2-dimensional planes
 defined by the abundances $Y_2$, $Y_4$, and $Y_7$, as functions of
 $x\equiv\log_{10}(\eta/10^{-10})$. The theoretical uncertainties are
 depicted as $1\sigma$ error ellipses at $x=0,\,0.1,\,0.2,\,\ldots,\,1$.
 The crosses indicate the two observational data sets (with $1\sigma$
 errors).}
\label{fig5}
\end{figure}

	Before performing the $\chi^2$ fit, it is useful to get an
idea of what one should expect by comparing the data with the
theoretical predictions at various values of $\eta$. Fig.~\ref{fig5}
shows the theoretical predictions for the abundances $Y_2$, $Y_4$, and
$Y_7$ in the three possible planes $(Y_i,\,Y_j)$, for representative
values of $x\equiv\log_{10}(\eta/10^{-10})$. The corresponding
$1\sigma$ error ellipses show clearly the size and the correlation of
the ``theoretical'' errors. The observational data sets ``A'' and
``B'' are also indicated on the figure, as crosses with $1\sigma$
error bars. Clearly the former prefers $\eta\sim2\times10^{-10}$ while
the latter favours $\eta\sim{}$(4--5)${}\times10^{-10}$.

\begin{figure}[tbh]
\epsfysize21truecm
\epsfbox{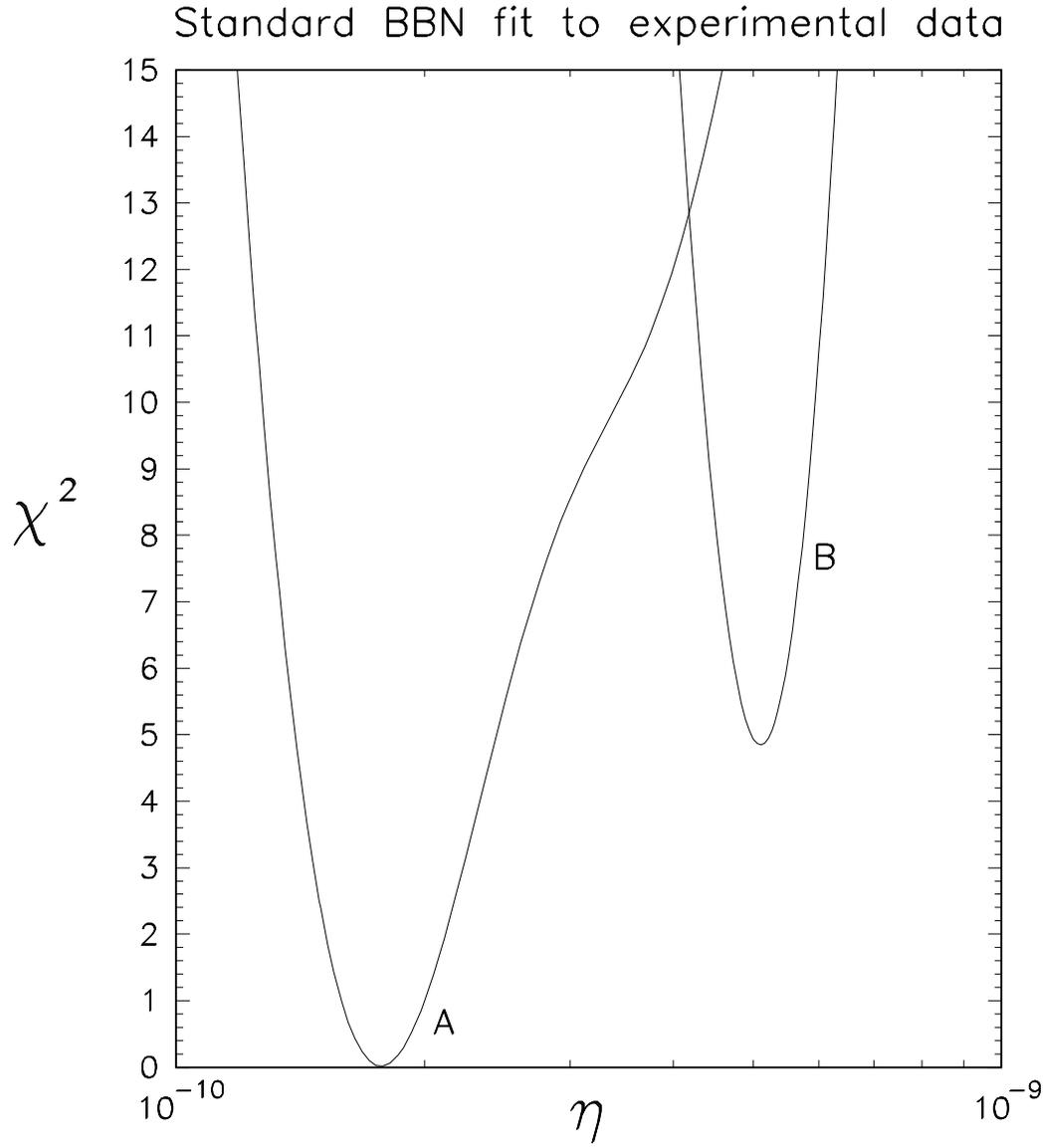}
\caption{Our $\chi^2$ fit to the data sets A and B, including
 observational and (correlated) theoretical errors.}
\label{fig6}
\end{figure}

	A more precise estimate of the likely range of $\eta$ is
obtained, as anticipated, through a $\chi^2$ fit. The results are
shown in Fig.~\ref{fig6}.  The value of $\chi^2_{\rm min}$ is almost
zero for the fit to data set ``A'', indicating very good agreement
between theory and observations, while it is somewhat larger for data
set ``B''. Note that the characteristic minimum in the $^7$Li curve at
$\eta\approx2.6\times10^{-10}$ (see bottom panel of Fig.~\ref{fig1})
allows its measured abundance to be compatible with both high D/low
$^4$He (data set ``A'') or low D/high $^4$He (data set ``B''). The
95\% C.L.\ ranges allowed by each of the two data sets, as obtained by
cutting the curves at $\Delta\chi^2=\chi^2-\chi^2_{\rm min}= 3.84$,
are:
\begin{eqnarray}
{\rm Data\ set\ A:\ }\ \eta & = & 1.78^{+0.54}_{-0.34} \times 10^{-10}\ ,\\
{\rm Data\ set\ B:\ }\ \eta & = & 5.13^{+0.72}_{-0.66} \times 10^{-10}\ .
\end{eqnarray}
The range for case ``A'' agrees very well with the 95\% C.L.\ range
estimated in Ref.\cite{Ol97c} with the same inputs but with a different
method (Monte Carlo + maximum likelihood). Of course, the
incompatibility between the above two ranges of $\eta$ reflects the
incompatibility between the input abundance data within their stated
errors.


\section{Role of different reactions in light element nucleosynthesis}

	The role of the different nuclear reactions rates listed in
Table~II in the synthesis of the light elements can be studied by
``perturbing'' the values of the input reaction rates and observing
their effect on the predicted abundances. More precisely, one can
study the contribution to the total uncertainty $\sigma_i$ of $Y_i$
induced by a $+1\sigma$ shift of $R_k$:
\begin{equation}
R_k \to R_k+\Delta\,R_k\ \Longrightarrow\ Y_i \to Y_i + \delta\,Y_i\ .
\end{equation}
Within our approach, this can be done very easily using Eq.~(3), with
the $\Delta\,R_k$'s from Table~II. Of course, the results depend on
the value chosen for $\eta$. To illustrate various trends, we choose
the best-fit values $\eta=1.78\times10^{-10}$ and
$\eta=5.13\times10^{-10}$, corresponding to data sets ``A'' and ``B''
respectively.

\begin{figure}[tbh]
\epsfysize20truecm
\epsfbox{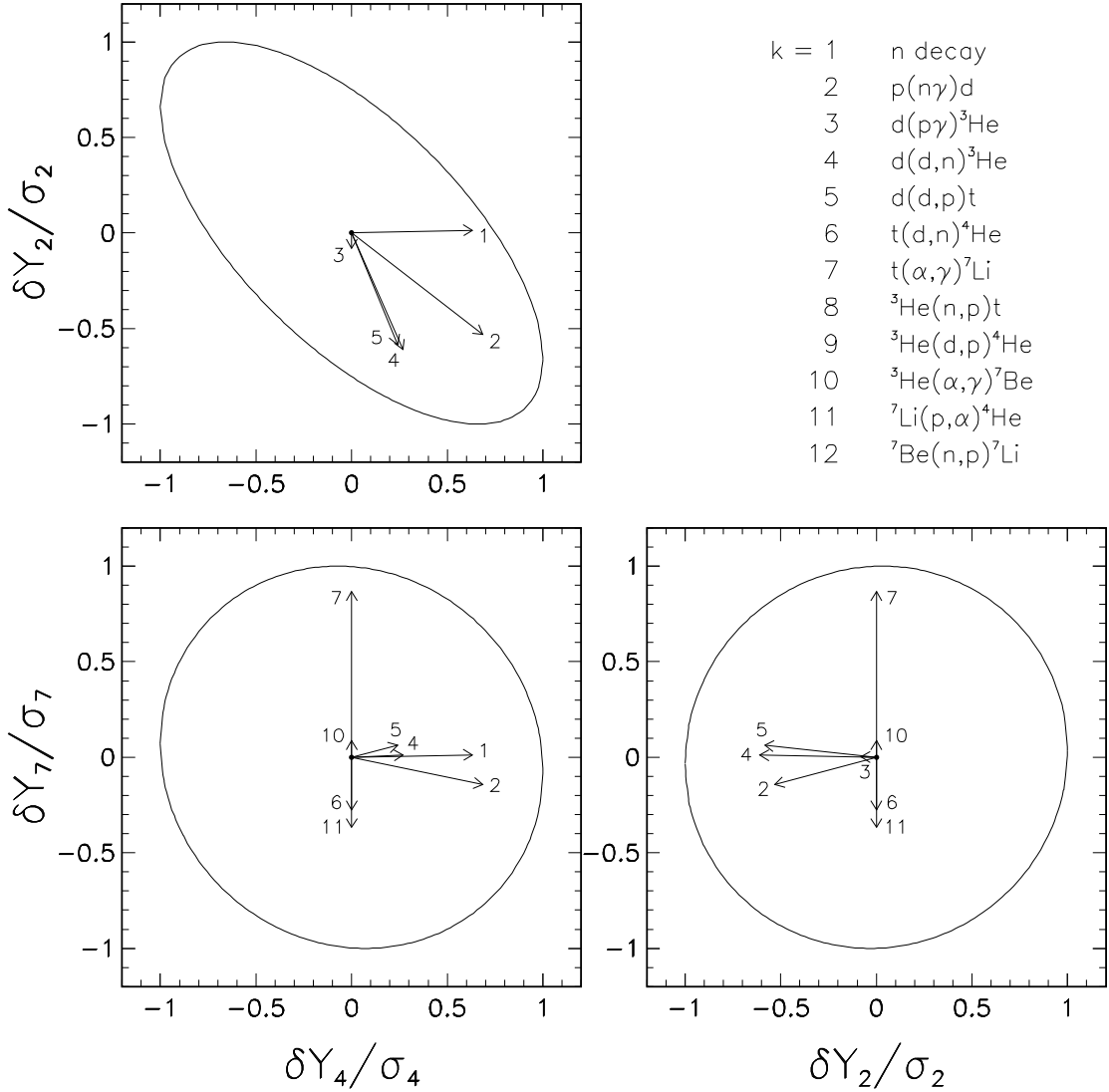}
\caption{Individual contributions of different reaction rates $R_k$ to
 the uncertainties in $Y_2$, $Y_4$, and $Y_7$, normalized to the
 corresponding total errors $\sigma_2$, $\sigma_4$, and $\sigma_7$,
 for $\eta=1.78\times10^{-10}$. Each arrow corresponds to the shift
 $\delta\,Y_i$ induced by a $+1\sigma$ shift of $R_k$. Some small
 error components have not been plotted.}
\label{fig7}
\end{figure}

\begin{figure}[tbh]
\epsfysize20truecm
\epsfbox{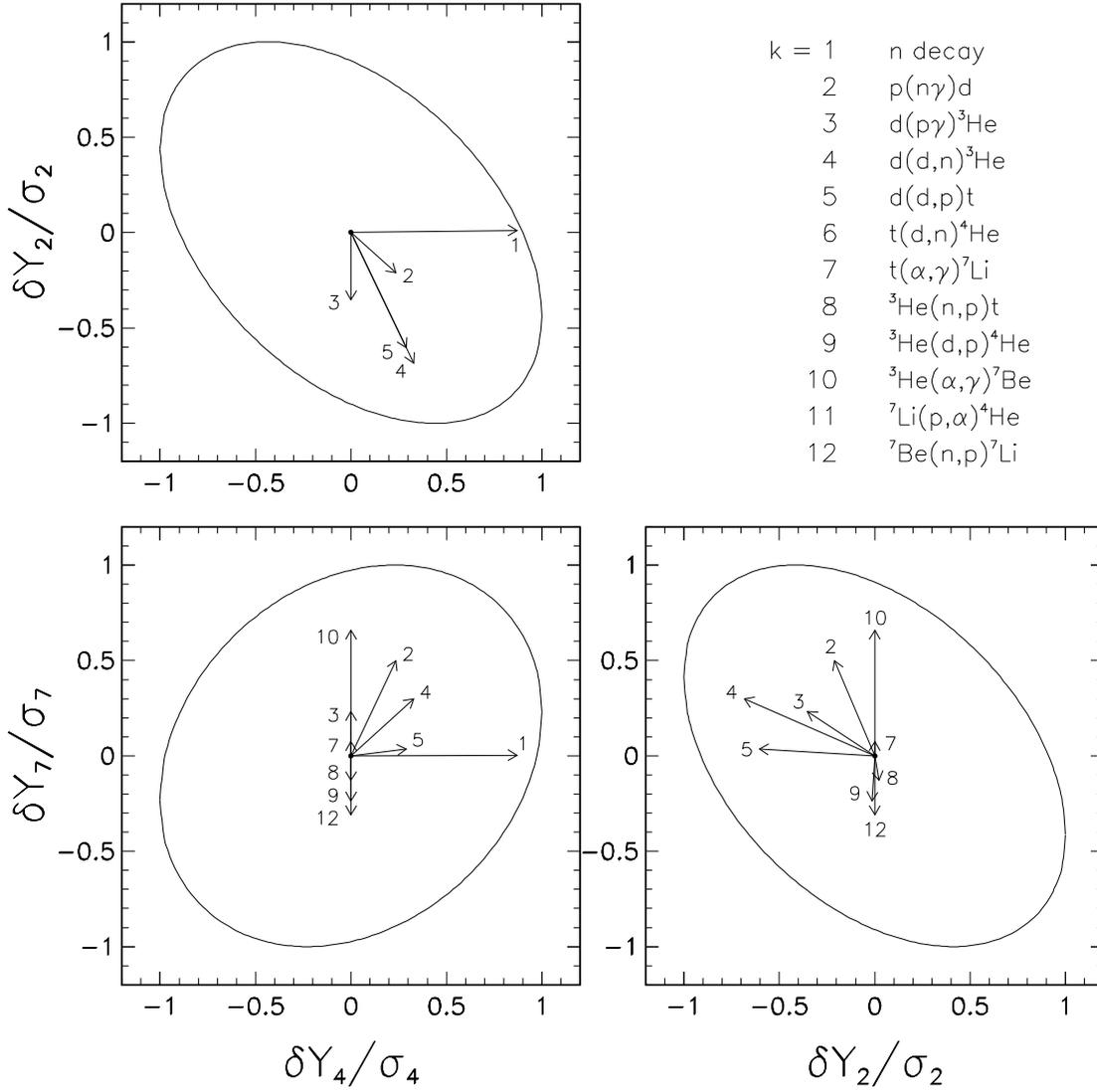}
\caption{Same as Fig.~\protect\ref{fig7}, but for $\eta=5.13\times10^{-10}$.}
\label{fig8}
\end{figure}

	Figs.~\ref{fig7} and \ref{fig8} show the deviations
$\delta\,Y_i$ (normalized to the total error $\sigma_i$) induced by
$+1\sigma$ shifts in the $R_k$'s, plotted in the same set of planes as
used for Fig.~\ref{fig5}. The $1\sigma$ error ellipses shown in these
figures are obtained by combining the deviation vectors $\delta
Y_i/\sigma_i$ in an uncorrelated manner. Several interesting
conclusions can be drawn from this exercise. As expected, the
uncertainty in the weak interaction rate $R_1$ has the greatest impact
on $Y_4$ for the high value of $\eta$ (Fig.~\ref{fig8}), since
essentially all neutrons end up being bound in $^4$He. However at the
lower value of $\eta$ (Fig.~\ref{fig7}), the uncertainty in $R_2$ ---
the ``deuterium bottleneck'' --- plays an equally important role as
$R_1$ in determining $Y_4$ because nuclear burning is less complete
here than at high $\eta$. Similarly with reference to the reaction
rates $R_7$, $R_{10}-R_{12}$ which synthesize $^7$Li, at low $\eta$ it
is the competition between $R_7$ and $R_{11}$ which largely determines
$Y_7$, while at high $\eta$ it is the competition between $R_{10}$ and
$R_{12}$. The anticorrelation between $Y_4$ and $Y_2$ is driven mainly
by $R_2$ at low $\eta$ and, to a lesser extent, by $R_4$ and $R_5$,
while the reverse is the case at high $\eta$. The anticorrelation
between $Y_4$ and $Y_7$ at low $\eta$ is also basically driven by
$R_2$, while the correlation at high $\eta$ is due to both $R_2$ and
$R_4$. Thus we have a direct visual basis for assessing in what
direction the output abundances $Y_i$ are pulled by possible changes
in the input cross sections $R_k$.


\section{Conclusions}

	We have shown that a simple method based on linear error
propagation allows us to quantify the uncertainties associated with
the elemental abundances expected from big bang nucleosynthesis, in
excellent agreement with the results obtained from Monte Carlo
simulations. This method makes transparent which nuclear reaction rate
is mainly responsible for the uncertainty in the abundance of a given
element. If determinations of the primordial abundances improve to the
point where the observational errors become smaller than the
theoretical uncertainties (say for $^7$Li), this will enable attention
to be focussed on the particular reaction rate whose value needs to be
experimentally better known.

	We have also demonstrated that for standard BBN, our method
enables the use of simple $\chi^2$ statistics to obtain the best-fit
value of $\eta$ from the comparison of theory and observations. At
present there are conflicting claims regarding the primordial
abundances of, particularly, D and $^4$He, and different choices of
input data sets imply values of $\eta$ differing by a factor of
$\sim\,3$. However this quantity can also be determined through
measurements of the angular anisotropy of the cosmic microwave
background (CMB) on small angular scales. Within a decade the
forthcoming all-sky surveyors MAP and PLANCK are expected to pinpoint
the nucleon density to within $\sim5\%$ \cite{cmb}. Such measurements
probe the acoustic oscillations of the coupled photon-matter plasma at
the (re)combination epoch and will thus provide an independent check
of BBN, assuming $\eta$ did not change significantly between the two
epochs.%
\footnote{New physics beyond the Standard Model can change $\eta$,
 e.g. by increasing the photon number through massive particle decay
 \cite{decay} or, more exotically, by {\em decreasing} the photon
 number through photon mixing with a shadow sector
 \cite{Ba91}. However such possibilities are strongly constrained by
 the absence of distortions in the Planck spectrum of the CMB
 \cite{El92} and also, in the latter case, by the absence of Sakharov
 oscillations in the power spectrum of large-scale structure
 \cite{Bi97}.}
Nevertheless precise measurements of light element abundances,
particularly $^4$He, are still crucial because they provide a unique
probe of physical conditions, in particular the expansion rate at the
BBN epoch. To illustrate, if $\eta$ was determined by the CMB
measurements to be $\approx\,2\times10^{-10}$ (consistent with data
set ``A''), but the abundance of $^4$He was established to be actually
closer to its higher value of $\approx\,24\%$ in data set ``B'', this
would be a strong indication that the expansion rate during BBN was
higher than in the standard case with $N_\nu=3$ neutrinos. Although
the number of $SU(2)$ doublet neutrinos is indeed 3, there are many
light particles expected in extensions of the Standard Model,
e.g. singlet neutrinos, which can speed up the expansion rate during
nucleosynthesis \cite{Sa96}. The generalization of our method to such
non-standard cases is straightforward and we intend to present these
results in a future publication \cite{us}. It is clear that BBN
analyses will continue to be important in this regard for both
particle physics and cosmology.

\acknowledgments

	E.L.\ thanks the Department\ of Physics at Oxford University
for hospitality during the early stages of this work. S.S.\ and F.V.\
thank the organizers and participants in the International Workshop on
{\em Synthesis of Light Nuclei in the Early Universe} held at ECT,
Trento in June 1997 for useful discussions. We thank Geza Gyuk and Rob
Lopez for very helpful feedback. This work was supported by the EC
Theoretical Astroparticle Network CHRX-CT93-0120 (DG12 COMA).


\begin{table}
\caption{The four light elemental abundances $Y_i$ considered in this
work. Alternative symbols used in the literature are indicated in
parentheses.}
\begin{tabular}{ccc}
         Symbol & ... or	&    Definition 		\\
\tableline
	$Y_4$  & ($Y_p$) 	&  $^4$He\ mass fraction	\\
	$Y_2$  & ($y_{2p}$)	&  D/H 	   \ \	(by number)	\\
	$Y_3$  & ($y_{3p}$)	&  $^3$He/H\ \	(by number)	\\
	$Y_7$  & ($y_{7p}$)	&  $^7$Li/H\ \ 	(by number)	
\end{tabular}
\end{table}

\begin{table}
\caption{The BBN reaction rates $R_k$ and their $1\sigma$
uncertainties $\pm \Delta\,R_k$ adopted in this work. The numbering
follows Ref.\protect\cite{Ka92} while the reference ``unit'' values
$(R_k\equiv 1)$ correspond to the rates in Ref.\protect\cite{Sm93}.}
\begin{tabular}{cccc}
$k$ 	&    Reaction 			 & $R_k$ & $\pm \Delta\,R_k$  \\
\tableline
1	& n~$\to$~p~e~$\bar\nu_e$	 & 0.9979 & $\pm 0.0021$\\
2	& p(n,$\gamma$)d		 & 1      & $\pm 0.07 $\\
3	& d(p,$\gamma$)$^3$He		 & 1      & $\pm 0.10 $\\
4	& d(d,n)$^3$He			 & 1      & $\pm 0.10 $\\
5	& d(d,p)t			 & 1      & $\pm 0.10 $\\
6	& t(d,n)$^4$He			 & 1      & $\pm 0.08 $\\
7	& t($\alpha$,$\gamma$)$^7$Li	 & 1      & $\pm 0.26 $\\
8	& $^3$He(n,p)t			 & 1      & $\pm 0.10 $\\
9	& $^3$He(d,p)$^4$He		 & 1      & $\pm 0.08 $\\
10	& $^3$He($\alpha$,$\gamma$)$^7$Be& 1      & $\pm 0.16 $\\
11	& $^7$Li(p,$\alpha$)$^4$He	 & 1      & $\pm 0.08 $\\
12	& $^7$Be(n,p)$^7$Li		 & 1      & $\pm 0.09 $
\end{tabular}
\end{table}

\begin{table}
\caption{Polynomial fit to the central value of the elemental
abundances, $Y_i=a_0 + a_1 x + a_2 x^2 + a_3 x^3 + a_4 x^4
+a_5 x^5$, with $x\equiv\log_{10}(\eta/10^{-10})$ in the range
0--1. The abundances were obtained using the BBN computer code
\protect\cite{Ka92} with the input $R_k$'s as in Table~II. The value
of $Y_4$ has been corrected using the prescription of
Ref.~\protect\cite{Sa96}. The accuracy of the fit is better than
$1/25$ of the total theoretical uncertainty for each $Y_i$.}
\begin{tabular}{ccccccc}
    	       &  $a_0$  &  $a_1$  &  $a_2$  &  $a_3$  &  $a_4$  &  $a_5$  \\
\tableline
$Y_2\times10^3$&$+0.4808$&$-1.8112$&$+3.2564$&$-3.3525$&$+1.8834$&$-0.4458$\\
$Y_3\times10^5$&$+3.4308$&$-6.1701$&$+8.1311$&$-9.7612$&$+7.7018$&$-2.5244$\\
$Y_4\times10^1$&$+2.2305$&$+0.5479$&$-0.6050$&$+0.6261$&$-0.3713$&$+0.0949$\\
$Y_7\times10^9$&$+0.5369$&$-2.8036$&$+7.6983$&$-12.571$&$+12.085$&$-3.8632$
\end{tabular}
\end{table}

\begin{table}
\caption{Polynomial fits to the logarithmic derivatives,
$\lambda_{ik}(x)$ = $a_0 + a_1 x + a_2 x^2 + a_3 x^3 + a_4 x^4 +a_5
x^5$, as functions of $x\equiv\log_{10}(\eta/10^{-10})\in[0,1]$ (see
Fig.~\protect\ref{fig2}). In most cases, polynomials of degree $<5$ provide
sufficently accurate fits.  Only non-negligible logarithmic
derivatives are tabulated.}
\begin{tabular}{cccccccc}
    	& $k$   & $a_0$ & $a_1$ & $a_2$ & $a_3$ & $a_4$ & $a_5$ \\
\tableline
$\lambda_{2k}$	
	&  1  &$+0.7130$&$-0.7964$&$+0.4577$&$+0.0914$&$      0$&$      0$\\
	&  2  &$-0.7025$&$+0.2611$&$+1.2008$&$-0.8934$&$      0$&$      0$\\
	&  3  &$-0.0189$&$-0.1879$&$+0.2502$&$-0.6806$&$      0$&$      0$\\
	&  4  &$-0.4228$&$-0.1698$&$+0.0207$&$+0.0247$&$      0$&$      0$\\
	&  5  &$-0.4138$&$-0.1477$&$+0.1103$&$+0.0010$&$      0$&$      0$\\
	&  8  &$-0.0073$&$-0.0003$&$+0.0801$&$-0.0416$&$      0$&$      0$\\
	&  9  &$-0.0011$&$-0.0348$&$+0.0592$&$-0.0511$&$      0$&$      0$\\
\tableline
$\lambda_{3k}$	
	&  1  &$+0.0940$&$+0.1892$&$-0.1484$&$-0.0199$&$      0$&$      0$\\
	&  2  &$+0.0981$&$+0.9948$&$-3.1667$&$+3.4108$&$-1.2845$&$      0$\\
	&  3  &$+0.0610$&$+0.1640$&$+0.5368$&$-0.2605$&$      0$&$      0$\\
	&  4  &$+0.3050$&$+0.0805$&$-0.4208$&$+0.1555$&$      0$&$      0$\\
	&  5  &$-0.5118$&$+0.1274$&$+0.4081$&$-0.2106$&$      0$&$      0$\\
	&  6  &$-0.0327$&$+0.0829$&$-0.0939$&$+0.0362$&$      0$&$      0$\\
	&  8  &$-0.5580$&$-0.0287$&$+1.3574$&$-0.8735$&$      0$&$      0$\\
	&  9  &$-0.1080$&$-0.5089$&$-1.0157$&$+0.8163$&$      0$&$      0$\\
\tableline
$\lambda_{4k}$	
	&  1  &$+0.8138$&$-0.1465$&$+0.0408$&$      0$&$      0$&$      0$\\
	&  2  &$+0.0610$&$-0.1962$&$+0.2416$&$-0.1049$&$      0$&$      0$\\
	&  4  &$+0.0082$&$-0.0058$&$+0.0034$&$      0$&$      0$&$      0$\\
	&  5  &$+0.0075$&$-0.0058$&$+0.0034$&$      0$&$      0$&$      0$\\
\tableline
$\lambda_{7k}$	
	&  1  &$+1.9638$&$+1.8520$&$-19.721$&$+27.542$&$-11.041$&$      0$\\
	&  2  &$-0.9214$&$-1.6472$&$+5.6187$&$+54.059$&$-112.80$&$+56.426$\\
	&  3  &$+0.0500$&$-1.0433$&$+4.1384$&$-2.4441$&$      0$&$      0$\\
	&  4  &$+0.1734$&$-2.7428$&$+11.209$&$-10.736$&$+2.5263$&$      0$\\
	&  5  &$+0.1837$&$-0.0875$&$+0.0158$&$-0.1350$&$      0$&$      0$\\
	&  6  &$-0.9877$&$-0.8168$&$+6.1555$&$-4.4380$&$      0$&$      0$\\
	&  7  &$+0.9644$&$+0.6888$&$-5.6151$&$+4.0284$&$      0$&$      0$\\
	&  8  &$+0.0529$&$+0.6318$&$-3.3995$&$+2.5754$&$      0$&$      0$\\
	&  9  &$-0.0764$&$+0.3017$&$-2.9632$&$+1.9311$&$      0$&$      0$\\
	&  10 &$+0.0690$&$-1.5360$&$+9.5808$&$-10.168$&$+2.9807$&$      0$\\
	&  11 &$-1.4095$&$-0.3543$&$+6.4780$&$-4.7956$&$      0$&$      0$\\
	&  12 &$-0.0043$&$-0.3779$&$+7.7358$&$-42.390$&$+61.401$&$-26.778$  
\end{tabular}
\end{table}




\eject
\end{document}